\documentclass[sigconf]{acmart}

\AtBeginDocument{%
  }

\setcopyright{acmcopyright}
\copyrightyear{2018}
\acmYear{2018}
\acmDOI{XXXXXXX.XXXXXXX}

\acmConference[Conference acronym 'XX]{Make sure to enter the correct
  conference title from your rights confirmation emai}{June 03--05,
  2018}{Woodstock, NY}
\acmPrice{15.00}
\acmISBN{978-1-4503-XXXX-X/18/06}

\usepackage{subfigure}
\usepackage{makecell}

\usepackage{enumitem}
\setitemize{noitemsep,topsep=0pt,parsep=0pt,partopsep=0pt}

\usepackage[english]{babel}
\addto\extrasenglish{%
}

\usepackage{forloop}

\newsavebox\MyBreakChar%
\sbox\MyBreakChar{}% char to display the break after non char
\newsavebox\MySpaceBreakChar%
\makeatletter%
\newcommand*{\BreakableChar}[1][\MyBreakChar]{%
  \leavevmode%
  \discretionary{\usebox#1}{}{}%
}%
\makeatother

\newcounter{index}%
\newcommand{\AddBreakableChars}[1]{%
  \StrLen{#1 }[\stringLength]%
  \forloop[1]{index}{1}{\value{index}<\stringLength}{%
    \StrChar{#1}{\value{index}}[\currentLetter]%
    \IfStrEqCase{\currentLetter}{%
        {*}{\currentLetter\BreakableChar[\MyBreakChar]}%
        {/}{\currentLetter\BreakableChar[\MyBreakChar]}%
        {+}{\currentLetter\BreakableChar[\MyBreakChar]}%
        {\&}{\currentLetter\BreakableChar[\MyBreakChar]}%
    }[\currentLetter]%
  }%
}%

\def\manualB{\raisebox{-0.15\height}{\includegraphics[height=1em]{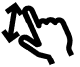}}\textbf{Manual}}

\def\guideB{\raisebox{-0.15\height}{\includegraphics[height=0.9em]{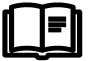}}\textbf{Compositional}}
\def\guide{\raisebox{-0.15\height}{\includegraphics[height=0.9em]{guide_B.pdf}}Compositional}

\def\manualT{\textsc{Manual}}
\def\guideT{\textsc{Compositional}}

\begin{document}

\title[]{Towards an Embodied Composition Framework for \\ Organizing Immersive Computational Notebooks}
% ICoN: Immersive Computational Notebook for Data Science with Embodied Data Transformation and Visualization

\author{Sungwon In}\orcid{0000-0002-5316-2922}
\affiliation{%
  \institution{Virginia Tech}
  \city{Blacksburg}
  \country{United States}}
\email{sungwoni@vt.edu}

\author{Eric Krokos}\orcid{0000-0003-1350-5297}
\affiliation{%
  \institution{Department of Defense}
  %\city{Blacksburg}
  \country{United States}}
% \email{ericpkrokos@gmail.com}

\author{Kirsten Whitley}\orcid{0000-0003-1356-326X}
\affiliation{%
  \institution{Department of Defense}
  %\city{Blacksburg}
  \country{United States}}
% \email{visual.tycho@gmail.com}

\author{Chris North}\orcid{0000-0002-8786-7103}
\affiliation{%
  \institution{Virginia Tech}
  \city{Blacksburg}
  \country{United States}}
\email{north@cs.vt.edu}

\author{Yalong Yang}\orcid{0000-0001-9414-9911}
\affiliation{%
  \institution{Georgia Tech}
  \city{Atlanta}
  \country{United States}}
\email{yalong.yang@gatech.edu}

% \authororcid{Kirsten\ Whitley}{0000-0003-1356-326X}
% \authororcid{Eric\ Krokos}{0000-0003-1350-5297}

\renewcommand{\shortauthors}{Trovato et al.}


\begin{abstract}
As immersive technologies evolve, immersive computational notebooks offer new opportunities for interacting with code, data, and outputs. 
However, scaling these environments remains a challenge, particularly when analysts manually arrange large numbers of cells to maintain both execution logic and visual coherence.
To address this, we introduce an embodied composition framework, facilitating organizational processes in the context of immersive computational notebooks.
To evaluate the effectiveness of the embodied composition framework, we conducted a controlled user study comparing manual and embodied composition frameworks in an organizational process. 
The results show that embodied composition frameworks significantly reduced user effort and decreased completion time. 
However, the design of the triggering mechanism requires further refinement.
Our findings highlight the potential of embodied composition frameworks to enhance the scalability of the organizational process in immersive computational notebooks.
\end{abstract}




\begin{teaserfigure}
  \centering
  \includegraphics[width=0.8\textwidth]{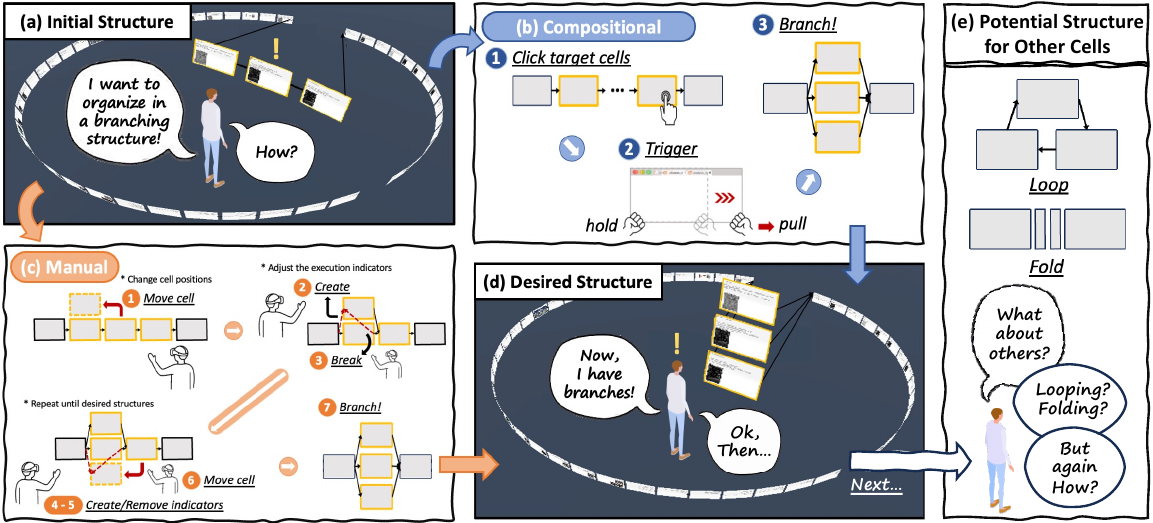}
  \vspace{0.4em}
  \caption{Comparison of how users organize cells in immersive computational notebooks. Initially, (a) the user aims to organize the linear arrangement of cells into a (d) desired structure. To organize efficiently, the user considers two approaches: (b) the embodied Composition framework and (c) Manual organization. (e) Users then wonder how to organize other cells.}
  \label{fig:teaser}
  \vspace{1em}
\end{teaserfigure}

\maketitle

\section{Introduction}

% Introduce challenges with scalability
Computational notebooks, such as Jupyter Notebook~\cite{JupyterNotebook}, are widely adopted by data analysts for their ability to integrate code, data, and visualizations into a sequence of cells.
However, as data grows in complexity, the increasing number of cells often leads to cluttered workspaces, making it difficult to clearly convey analytical intent~\cite{head2019managing}.
While analysts may attempt to organize their notebooks, limited workspace in desktop settings remains challenging~\cite{chattopadhyay2020s}.
% due to the limited screen size and two-dimensional layouts~\cite{chattopadhyay2020s}.}

Meanwhile, the use of immersive technologies in data analytics has shown great potential~\cite{ens2021grand, marriott2018immersive, yang2020tilt}. 
Notably, extending computational notebooks into immersive environments facilitates workspace management by spatially organizing cells that align with mental models and analytical goals, all within a large 3D workspace~\cite{in2024evaluating}. 
Although such intuitive organization helps analysts better understand complex workflows~\cite{andrews2010space}, immersive computational notebooks present significant challenges in organizing them.
Specifically, data analysts can organize one or two cells at a time, and manually manage their alignment~\cite{in2025exploring}.
While content organization in immersive environments has been widely studied~\cite{luo2025documents, davidson2024investigating}, execution orders that reflect the inherent logics between cells introduce an additional complexity in the organizational process.
Therefore, analysts must interact with cells plus associated execution orders one by one, which becomes effort-intensive~\cite{batch2019there, kraus_impact_2019}.
\textbf{Building on this, our primary goal is to reduce the effort required for organizational processes in immersive computational notebooks.}
% enabling analysts to focus more on their analytical tasks and derive deeper insights.

% Content organization in immersive environments has been widely studied, such as Luo et al. examined organization in the presence of real-world obstacles~\cite{luo2025documents}, and Davidson et al. explored the relationship between organizational strategies and sensemaking processes~\cite{davidson2024investigating}.
% Yet, prior efforts have largely focused on document-based tasks.  
% Immersive computational notebooks, however, introduce an additional layer of complexity in the organizational process: execution orders, which reflect the inherent logics between code cells.
% Therefore, analysts must interact with spatially distributed cells plus associated execution orders one by one manually, which becomes effort-intensive~\cite{batch2019there, kraus_impact_2019}.
% Most importantly, as the number of cells increases, these efforts grow significantly, highlighting a limitation in scaling immersive computational notebooks.
% \textbf{Building on this, our primary goal is to reduce the effort required for organizational processes in large-scale immersive computational notebooks, enabling analysts to focus more on their analytical tasks and derive deeper insights.}

Common practices for reducing the effort required in organizational processes can be observed in a desktop setup.
For instance, Miro~\cite{miro} and Figma~\cite{figma} support a composition framework that consists of predefined visual layouts, enabling users to rapidly initiate structured workflows with less effort, and facilitating clarity and consistency throughout the process~\cite{collins2018guidance}. 
Moreover, this composition framework has been shown to support scalability by offering reusable structures that simplify the organization of large analytical workspaces and reduce repetitive manual adjustments~\cite{miller2022augmenting}.

While the composition framework is promising, designing for immersive computational notebooks requires additional considerations.
The composition should account not only for the layout but also ensure that the execution order within the layout is clearly conveyed.
In addition, depth cues should also be considered, as users frequently put higher priority cells closer for quicker access~\cite{in2023table}.
Moreover, traditional WIMP (Windows, Icons, Menus, Pointer) interfaces are effective in 2D environments~\cite{jo_touchpivot_2017}; their use in immersive settings can disrupt the flow of interaction, as they require frequent switching between menus and target artifacts~\cite{in2024evaluating}.
In contrast, gesture-based embodied interaction in immersive contexts enables maintaining user engagement and supports more fluid and intuitive workflows~\cite{cordeil2019iatk}.

To this end, our work introduces an embodied composition framework for immersive computational notebooks, which consists of 10 unique structures, each parameterized by three key properties: execution order, layout, and analysis phase.
In addition, we designed gestural interaction to trigger them without distraction from external commands.
Furthermore, we supported flexible adjustments within existing organizations to adapt to evolving analysis.

To evaluate the embodied composition framework, we conducted a user study comparing it with a baseline condition simulating current practices in immersive computational notebooks, where analysts placed and adjusted only a few cells and their execution order at a time to organize cells.
Our findings revealed that the embodied composition framework allowed participants to build structures more quickly, as it removed the need to manipulate artifacts one by one.
However, we also identified that the triggering mechanism needs further improvement.

In summary, our main contributions are:
\begin{itemize}[leftmargin=2.3em]    
    \item We designed an embodied composition framework for immersive computational notebooks, a research prototype that aims to reduce the effort required in the organizational process.
    
    \item We conducted a user study to investigate how the presence of an embodied composition framework influences users' organizational process in an immersive computational notebook.
\end{itemize}

% !TEX root = ../main.tex

\begin{figure}
    \centering
    \includegraphics[width=0.85\columnwidth]{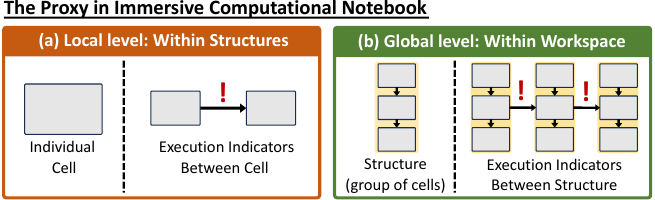}
    \vspace{0.3em}
    \caption{Proxies for organizing immersive computational notebooks. (a) local-level proxies within structures, involving individual cells and execution order between cells, and (b) global-level proxies within the workspace, involving structures and execution order between structures.}
    \label{fig:proxy}
\end{figure}

\begin{figure*}
    \centering
    \includegraphics[width=0.9\textwidth]{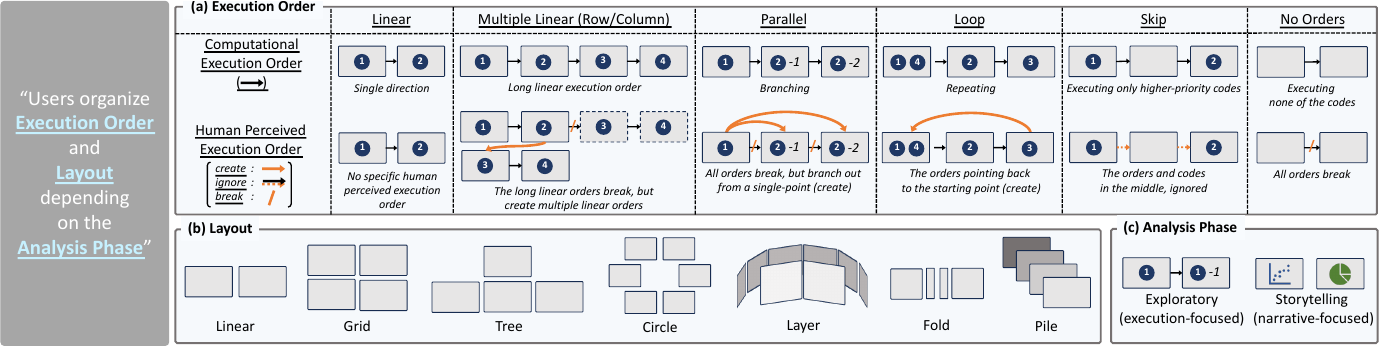}
    \vspace{0.4em}   
    \caption{Illustration of three components in organizing immersive computational notebooks: (a) Execution order, where the system visualizes execution linearly while humans perceive it differently; (b) Layout, describing how cells are positioned in 3D space; and (c) Analysis phase, representing types of use cases in computational notebooks.}
    \label{fig:design}
\end{figure*}

\section{Related Work} 
\label{sec:relwork}

% Talking about the executin order within computataional notebooks
\textbf{Computational notebooks.}
The computational notebook emerged from literate programming, aiming to integrate narrative, code, and visualization in a unified document~\cite{Kluyver2016jupyter}.
Most computational notebooks use a cell-based interface that displays code in a top-down linear sequence, reflecting the sequential nature of code execution.
While this linear structure enables the rapid testing of simple analyses~\cite{kery2018story}, it becomes limiting in supporting the various roles of computational notebooks---from exploratory analysis to storytelling---due to its lack of flexibility in repositioning cells~\cite{chattopadhyay2020s, tabard2008individual, wang2022stickyland}.
In response, prior work has explored non-linear notebook interfaces, such as Fork-it by Weinman et al., a system that enables analysts to test multiple analytical paths simultaneously~\cite{weinman2021fork}. 
While Fork-it supports practical analysis where execution logics are important, Kang et al. proposed ToonNote, which incorporates novel interaction techniques to highlight the visual narratives~\cite{kang2021toonnote}. 
Based on their successful explorations, recent work by In et al.~\cite{in2024evaluating} introduced immersive computational notebooks that extended notebook interfaces in 3D space. 
Immersive computational notebooks spatially arrange cells in a semi-circular layout across multiple floating windows, allowing users to explore across a wide field of view and flexibly organize within a large display space. 
While this shift enables analysts to move beyond linear constraints, it also raises important questions about how best to structure content at scale.

\textbf{Content organizations in immersive environment.}
Unlike traditional 2D interfaces, users in immersive spaces can position content in three dimensions, allowing for novel layout strategies that better reflect the structure and flow of analytical reasoning~\cite{knierim2020opportunities, ball_move_2007, liu2020design, satriadi2020maps}. 
For instance, Satriadi et al. found that cylindrical layout is effective in structuring hierarchical, multi-view geospatial datasets, which enables better separation of data layers~\cite{satriadi2020maps}.
In addition, Liu et al. further specify that a half-cylindrical structure was most effective for presenting multiple data visualizations, as it enhances access and readability within the user’s field of view~\cite{liu2020design}. 
% Furthermore, Luo et al.~\cite{luo2025documents} investigated how real-world environmental contexts influence the spatial arrangement of content in immersive environments.
Their findings revealed that the organization's strategies were decided by user positioning and physical surroundings, including the strategic use of depth, proximity, and anchoring to real-world objects.
In the context of computational notebooks, users preferred to display cells in a cylindrical layout and utilized spatial depth to position artifacts, either to convey hierarchical relationships or to prioritize certain content, resulting in layered structures~\cite{in2024evaluating}.
% In this work, we compile a comprehensive set of structures applicable to immersive computational notebooks.

% Talking about the template
\textbf{Assisted Organizational Composition.} 
External assistance in organizations aims to support users in structuring workflows by utilizing the composition of pre-defined structures~\cite{collins2018guidance, yilmaz2024comparison}.
Early work focused on rule-based layouts and alignment aids, which helped maintain clarity and consistency, particularly in domains such as design tools and collaborative whiteboards~\cite{plimmer2004shared, shen2025dashchat, george2005remote}. 
For instance, Figma~\cite{figma} enables quick initiation of projects using predefined templates that provide structured starting points, while Miro~\cite{miro} supports workflow visualization by offering templates with consistent spacing, grouping, and alignment. 
In the context of computational notebooks, most existing mechanisms are designed to support the progression of analysis, such as creating new branches~\cite{weinman2021fork}, but they rarely facilitate the existing organizations into desired structures.
While some systems have introduced features to reorganize notebook layouts, their focus is often on transforming notebooks into dashboards, narratives, or visual documents~\cite{kang2021toonnote}.
Moreover, depth, an additional dimension introduced in immersive environments, becomes a meaningful organizational axis.
However, the existing organizational composition did not thoroughly consider execution order within immersive environments.
% To address these gaps, we explore organizational assistance to facilitate the organizational process in a way that accounts for both execution order and spatial depth.

\section{Design Rationale}
\label{sec:design_goal}
We aim to support content organization in flow-based systems within immersive environments. 
In flow systems, users interact with spatial proxies across various levels of granularities~\cite{yu2016visflow, in2024evaluating}; positioning individual artifacts and flows at the local level, and managing a group of artifacts and their flows at the global level. 
In immersive computational notebooks, this translates to positioning individual cells and adjusting execution logic at the \textit{local level}, focusing within structures (see~\autoref{fig:proxy}-(a)) and relocating entire structures and their executions at the \textit{global level}, focusing within workspaces (see~\autoref{fig:proxy}-(b)).
Understanding this, we derived design principles for content organizations in immersive computational notebooks drawing on prior work in immersive analytics~\cite{cordeil2019iatk}, spatial interaction~\cite{yang_pattern_2022}, and computational notebook usage~\cite{kery2018story}.
These include: 1) support for diverse tasks, 2) intuitive input modality, 3) flexible override, and 4) support for scaling.
Based on these design principles, we detail proposed design goals.

% It is important to at least have a paragraph describing how you come up with those design goals. In our case, literature and HCI/vr/ar design guidelines?

\textbf{G1. Support organization both for execution and narrative-focused tasks.}
Computational notebooks support tasks from exploratory analysis to storytelling~\cite{lau2020design}.
Exploratory analysis focuses on execution order~\cite{chattopadhyay2020s}, whereas storytelling emphasizes visual narrative~\cite{kery2018story}. 
To support the wide range of uses in computational notebooks, the system should support organizations that either highlight execution orders or narratives clearly.

\begin{figure*}
    \centering
    \includegraphics[width=0.8\textwidth]{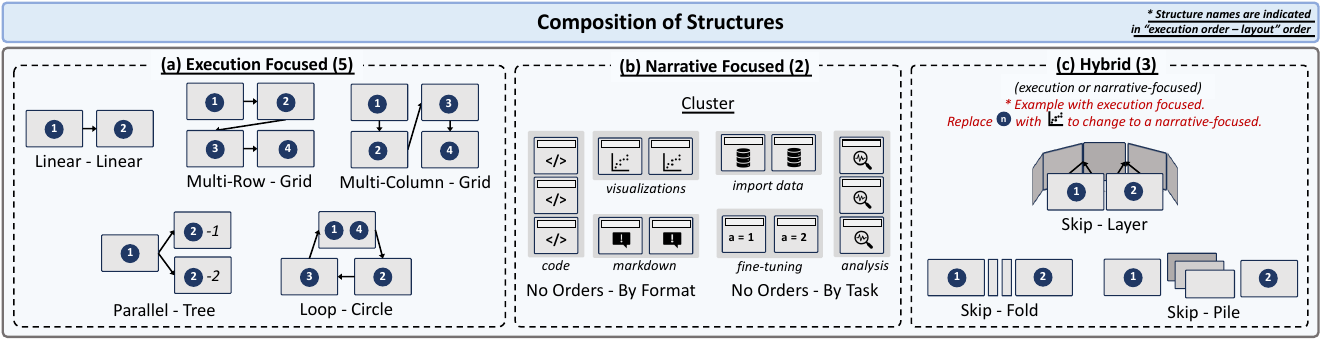}
    \vspace{0.4em}
    \caption{Proposed composition structures for the embodied composition framework. (a) execution-focused, where execution order is primary; (b) narrative-focused, where visualizations and results are more critical over execution order; and (c) hybrid, supporting both execution- and narrative-focused structures.}
    \label{fig:template}
\end{figure*}

\textbf{G2. Ensure an intuitive triggering mechanism.}
Embodied interactions in immersive environments offer fluid engagement through natural movements~\cite{reiske2023multi, yang_pattern_2022}. 
This form of interaction also extends to triggering mechanisms, where gestures serve as an input modality for executing operations.
Our goal is to leverage such gestures to design intuitive, embodied triggering mechanisms.

\textbf{G3. Support human intervention within the existing organizations.}
Data analysis is inherently dynamic; analysts explore multiple hypotheses, revisit earlier steps, and adjust their reasoning~\cite{rule2018exploration}---prompting changes in how they organize, prioritize, and structure their notebooks. 
While a composition framework helps with rapid setup, users may want refinements on existing organizations. 
Thus, systems should support flexible overrides and edits to the existing organizations.

\textbf{G4. Support effective global-level organizations.}
Analysts often adjust the position of groups of cells and refine their execution orders.
However, unlike interacting with an individual proxy, interacting with proxies as a group adds additional interaction cost~\cite{in2024evaluating}.
Furthermore, as the number of cells increases, manually interacting with them one by one becomes more laborious. 
Therefore, systems should allow users to move groups of cells along with their associated execution orders at once.

\section{Embodied Composition Framework for Immersive Computational Notebooks}
\label{sec:implementation}
We introduce the embodied composition framework, a prototype system designed to support users in building and organizing artifacts within immersive computational notebooks using embodied interactions.
The embodied composition framework consists of three main components: 1. compositions of structures, 2. triggering mechanisms, and 3. interactions with existing organizations.
Below, we describe each of these components in detail.

\subsection{Consideration for Compositions}
\label{sec:design_framework}
We began by reviewing the literature to identify key principles for designing a composition framework for organizing immersive computational notebooks. 
Specifically, we searched the ACM Digital Library, IEEE Xplore, and Google Scholar using keywords such as immersive analytics, computational notebooks, and spatial organization. 
Our review focused on the past decade of research in immersive environments and data analysis tools.
While literature has not fully addressed the organizational strategies that involve both execution order and spatial layout in immersive environments, it offers guiding principles. 
For instance, Luo et al. showed how users strategically manage spatial layouts~\cite{luo2025documents}, while Harden et al. explored how execution orders vary with analytical tasks in 2D settings~\cite{harden2022exploring}. 
Drawing from these insights, we propose a conceptual framing that involves three key design aspects: ``Users organize \textbf{\underline{Execution Order}} and \textbf{\underline{Layout}} depending on the \textbf{\underline{Analysis Phase}}'' (see~\autoref{fig:design}).
We describe each aspect below.

\textbf{Execution Order (6).}
Building on prior studies of organizational strategies and explorations aimed at enhancing analytic performance in computational notebooks~\cite{harden2022exploring, weinman2021fork, stoudt2021principles, reimann2023alternative, nakamaru2024multiverse}, we identified six commonly used execution orders. 
Among the identified execution orders, we additionally found that they present two perspectives. 
First, the computational execution order, which is system-enforced~\cite{rule2018exploration}, and the human-perceived execution order, which captures how analysts interpret the execution sequence~\cite{chattopadhyay2020s}.
We describe these six orders and how they differ from system logic and user perception.
\textit{(1) Linear:} The most common pattern in desktop settings like Jupyter Notebooks~\cite{JupyterNotebook}, where the human perceives the same as computational order. 
\textit{(2) Multiple Linear:} Analysts often organize into multiple rows or columns to break the long linear computational orders.
\textit{(3) Parallel:} Analysts create parallel execution paths from a single point, allowing multiple hypotheses to be executed simultaneously.
\textit{(4) Loop:} Analysts establish loops where the execution path circles back to prior steps.
\textit{(5) Skip:} Analysts selectively execute cells to optimize performance, such as computationally intensive cells.
\textit{(6) No-Order:} Analysts do not execute any cells, where the execution order is not the primary concern.

\textbf{Layout (7).}
Building on prior studies of spatial organization in immersive settings~\cite{in2024evaluating, luo2025documents, bach2015small, elmqvist2008melange}, we identified seven layouts commonly employed:
\textit{(1) Linear:} Artifacts are arranged sequentially in a single row or column.
\textit{(2) Grid:} Artifacts are distributed across multiple rows and columns.
\textit{(3) Tree:} Artifacts diverge or branch out from a parent artifact.
\textit{(4) Circle:} Artifacts are organized in a circular shape.
\textit{(5) Layer:} Artifacts are organized into smaller layers by leveraging spatial depth and separating artifacts from others.
\textit{(6) Fold:} All artifacts are collapsed.
\textit{(7) Pile:} Artifacts are stacked, with the first element remaining visible to represent the group.

\textbf{Analysis Phase (2).}
Prior work in immersive analytics highlights two primary phases in analytical workflows: exploratory and storytelling~\cite{ens2021grand}. 
Similarly, surveys of computational notebook usage reveal that notebooks are commonly used to support both phases~\cite{chattopadhyay2020s}. 
Therefore, the composition in immersive computational notebooks should be designed to accommodate the organizational needs of both phases:
\textit{(1) Exploratory analysis:} Analysts primarily focus on execution order to maintain logical consistency~\cite{rule2018exploration}.
\textit{(2) Storytelling:} Analysts prioritize highlighting key insights and group cells semantically, cluster outputs, and often ignore the execution flow to emphasize the narrative~\cite{kery2018story}.

\begin{figure*}
    \centering
    \includegraphics[width=0.9\textwidth]{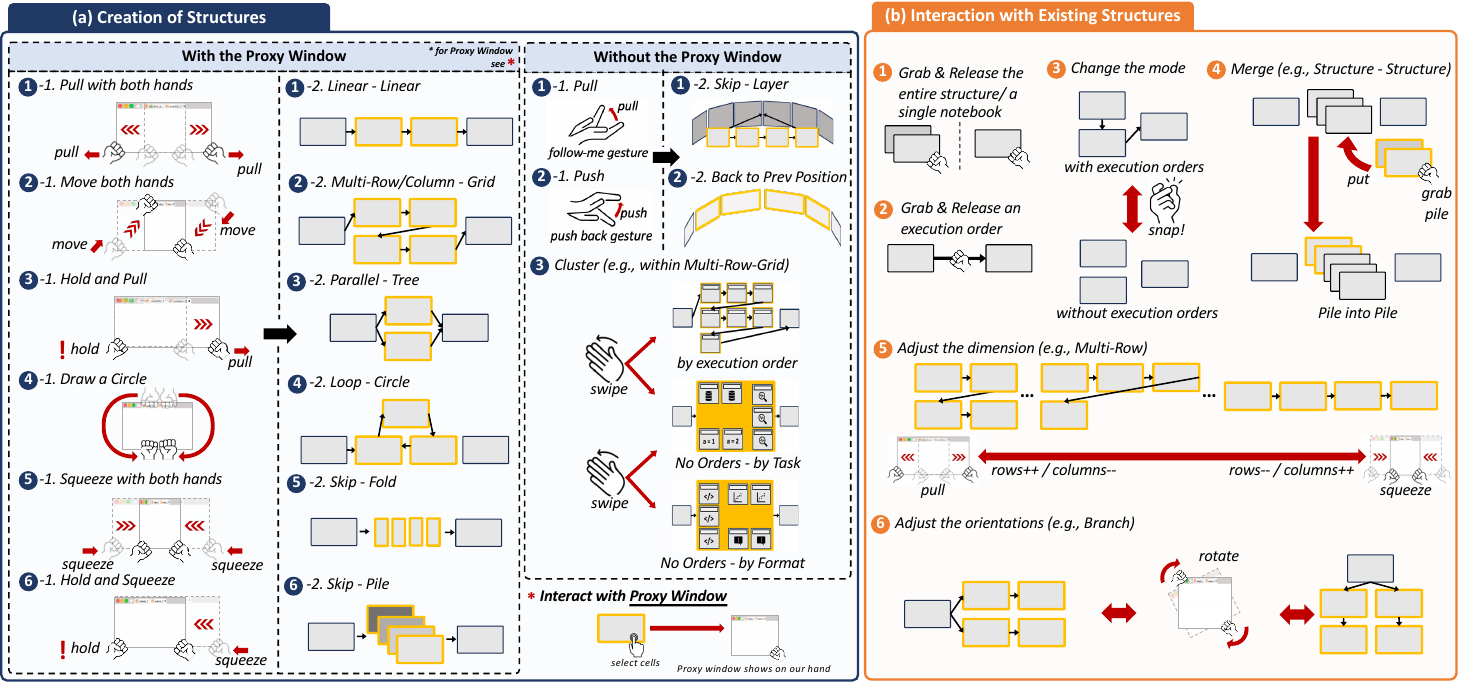}
    \vspace{0.5em}
    \caption{Illustrations of triggering mechanisms in the embodied composition framework. (a) creating desired structures, either through interaction with or without the proxy window; (b) interacting with existing structures.}
    \label{fig:interaction}
\end{figure*}

\subsection{Compositions of Structures}
\label{sec: structure}
When users manually organize artifacts, it often results in misaligned or inconsistent structures~\cite{in2023table}.
In contrast, predefined structures allow analysts to build well-organized structures for various types of tasks~\cite{kang2021toonnote, weinman2021fork}, aligning with our design goal of \textit{supporting organization for both execution-order and narrative-focused tasks (G1)}.
To this end, we support 10 unique predefined structures (see~\autoref{fig:template}) derived by considerations discussed in \autoref{sec:design_framework}.

\textbf{Execution Focused (5).}
In these structures, the primary focus is placed on maintaining coherent execution orders.
\textit{(1) Linear-Linear:} A single linear execution order arranged in a linear layout.
\textit{(2) Multi-Row-Grid:} Multiple linear execution orders organized into several rows, creating a grid layout.
\textit{(3) Multi-Column-Grid:} Multiple linear execution orders organized into several columns, creating a grid layout.
\textit{(4) Parallel-Tree:} Parallel execution orders that branch out from a single point, creating a tree layout.
\textit{(5) Loop-Circle:} An execution flow that forms a loop in a circular layout, supporting iterative or cyclic analysis processes.

\textbf{Narrative Focused (2).}
In these structures, the primary focus is on organizing cells to highlight key insights and enhance the clarity of the narrative, rather than adhering to execution order. 
\textit{(1) Cluster-by-Cell Format:} Cells are clustered by their content format (e.g., code, visualization, comments) to emphasize their functional roles. 
\textit{(2) Cluster-by-Task:} Cells are clustered according to distinct analytic tasks, highlighting high-level workflow structures.

\textbf{Hybrid (3).}
In these structures, the organization supports both coherent execution orders and narrative clarity.
\textit{(1) Skip-Layer:} Skips parts of the workflow by pulling specific cells closer in depth, creating a layered structure across different spatial levels.
\textit{(2) Skip-Fold:} Skips parts of the workflow by folding sequences, compressing the structure while preserving logical order.
\textit{(3) Skip-Pile:} Skips parts of the workflow by piling cells on top of each other, forming a stacked structure that allows flexible access across steps.

\subsection{Triggering mechanism}
\label{sec: ch5_trigger}
After we designed the composition of structures from \autoref{sec: structure}, we aim to maintain user immersion while organizing target cells and execution orders into one of the structures in the provided compositions.
Initially, we considered a WIMP approach, but this introduced high context-switching costs as it forces users to shift focus between target proxies and menu interfaces~\cite{in2023table}.
Moreover, accommodating numerous structure options would have resulted in either oversized menus or small buttons that demand high precision, problematic in immersive environments~\cite{siek2005fat}.
In contrast, embodied interaction enables seamless transitions between target proxies and actions~\cite{yang2020embodied}, supporting our design goal of \textit{fostering intuitive and fluid interaction (G2)}.
We therefore adopted gesture-based input to keep users engaged with the organization without being distracted by external commands. 
Gestures were iteratively refined based on collaborative design sessions and prior work~\cite{luo2025documents, zhu2024compositingvis, agarawala2006keepin, yang2020embodied}.
In the following sections, we describe how users can trigger the provided compositions of structures (see~\autoref{fig:interaction}).

\textbf{With the Proxy Window (6).}
In this interaction, all actions involved a ``grab'' gesture.
To enhance the sense of physically grasping an object, we provided visual feedback, the Proxy Window, which made users aware of their actions. 
Through the pilot study, we found that users were clearly aware of their actions and were not distracted by the presence of the Proxy Window.
The Proxy Window appears near the user’s hand when cells are selected (see the right bottom corner in Fig. 5-(a)) and supports structure creation through the following gestures:
\textit{(1) Linear-Linear:}  Pull outward the Proxy Window with both hands. This organizes the cells into a Linear-Linear structure.
\textit{(2) Multi-Linear-Grid:} Grab the Proxy Window and move their hands diagonally—imagining a top-to-bottom arrangement. This organizes the cells into a Multi-row-Grid structure (default). To switch to a column-based execution, users adjust the grid's orientation.
\textit{(3) Parallel-Tree:} Users select the first cell of each intended branch and pull outward the Proxy Window with one hand, mimicking a branching out from a single cell. This organizes the cells into a Parallel-Tree structure.
\textit{(4) Loop-Circle:} Grab the Proxy Window and move their hands in a circular motion, simulating a loop. This organizes the cells into a Loop-Circle structure.
\textit{(5) Skip-Fold:} Squeeze the Proxy Window with both hands. This organizes the cells into a Skip-Fold structure.
\textit{(6) Skip-Pile:} Squeeze the Proxy Window with one hand, imagining piling the cells on top of one another. This organizes the cells into a Skip-Pile structure.

\textbf{Without the Proxy Window (2).}
For specific interactions, a Proxy Window was unnecessary, as these gestures were distinguishable from others.
\textit{(1) Skip-Layer:} Users select either individual cells or an entire structure and perform a Follow-Me gesture or a Go-Away gesture. This organizes the cells into a skip execution order in spatial layers.
\textit{(2) No-Orders-Cluster:} Users place their hands near the existing structure and swipe to cluster the cells by format and task. Each swipe changes the clustering mode.

\subsection{Interaction with the existing organizations}
In addition, we introduced interactions with existing organizations to support evolving analytical workflows.
These interactions enable both the \textit{refinement of existing organizations (G3)} and \textit{facilitate global-level organizations (G4)}.
Similar to \autoref{sec: ch5_trigger}, we iteratively refined the gesture designs.
Below, we outline how the additional interactions can be performed. 

\textit{(1) Move along with a group of cells:}
After organizing cells and execution orders using the compositions of structures, a grabber appears at the structure’s center, only when users place their hands nearby to minimize occlusion. Users can grab this handle to move the entire structure and release it at a desired position.
\textit{(2) Grab single cell/execution indicator:}
Users can grab a single cell and release it away from the existing structure to remove it, then release it back into the structure to include.
Additionally, users can grab any of the execution indicators to adjust the execution order to better reflect the user's intentions.
\textit{(3) Change mode:}
Users can toggle the visibility of the execution indicator by selecting cells and performing a snapping action.
\textit{(4) Merge:}
Users can grab one structure and release it into another existing structure. 
The inserted structure is then transformed into the same format as the target structure.
\textit{(5) Adjust the dimensions:}
Users can adjust the number of rows or columns by squeezing or unsqueezing the Proxy Window after selecting cells in a Multiple-Row/Column-Grid structure. 
\textit{(6) Adjust the orientations:}
Users can change the orientation of Multiple-Row/Column-Grid and Parallel-Tree structures to either a horizontal or vertical manner by selecting cells and rotating the Proxy Window.

\section{User Study}
\label{sec:user_study}
We compare the embodied composition framework, where the system assists participants in organizations, with manual organization, where participants could organize only a few cells and execution orders at a time, and had to manually adjust alignments.
Furthermore, to investigate the effectiveness of the embodied composition framework, we simulated a large-scale data analysis scenario.
In summary, we address the following four research questions:

\textbf{RQ1. Does the embodied composition framework help analysts organize more quickly and with less effort?}
We aim to determine whether the embodied composition framework enables users to build structures more quickly and with minimal effort.

\textbf{RQ2. Do users continue to abandon organizational refinement when using the embodied composition framework?} 
As the number of artifacts increases, users frequently abandon organizational refinements when interacting with them one by one manually \cite{in2023table}. This often leads to inconsistent spacing, misalignments, and poorly placed execution indicators. Therefore, we aim to assess whether the embodied composition framework mitigates this tendency and results in more coherent final organizations.

\textbf{RQ3. Does the embodied composition framework reduce the variation in organizational strategies compared to manual organization?} In an assistance system (e.g., LLM recommendation), users often heavily follow the recommendations, leading to similar reasoning across individuals~\cite{vasconcelos2023explanations}. We anticipate that a similar pattern may emerge with the embodied composition framework, where assistance is introduced into the organizational process. Therefore, we aim to compare the organizational strategies of manual organization and the embodied composition framework.

\textbf{RQ4. Are the provided gesture-based triggering mechanisms intuitive and well-aligned with user intentions?}   
While the gestures were grounded in common interaction behaviors~\cite{luo2025documents}, mismatches may still occur. Therefore, we evaluate whether users found gestures reflective of their intended actions.

\subsection{Study Conditions and Data}
\label{sec:user_study_conditions}
To examine the effectiveness of the embodied composition framework, we conducted a user study comparing two conditions: with and without the embodied composition framework. 
We refer to the condition without the framework as \manualT{} and the condition with the framework as \guideT{}. 
The details of each condition are as follows:

\manualB{}: Participants used the same interaction methods from existing immersive computational notebooks. 
They could interact with only one or two cells or execution indicators at a time (grabbing one with each hand) to organize them into desired structures. Also, they had to manually adjust spacing and alignment.
    
\guideB{}: Participants had the same interactions as in the \manualT{}, but could transform multiple cells and execution order into desired structures with a single interaction (e.g., gesture).
While participants could still adjust positions, execution indicators, spacing, and alignment within transformed structures, the composition ensured that cells and execution indicators were placed consistently and clearly.

Regarding the data, we explored several configurations to determine the number of cells. 
While a larger set (e.g., 75 or more) could better highlight organizational differences between \manualT{} and \guideT{}, pilot testing revealed excessive completion times. 
However, using 20 code cells, as in prior immersive computational notebook studies~\cite{in2025exploring}, made the organizational task too quick to complete.
We gradually increased the number of cells in pilot testing and found that 50 provided a reasonable balance, demanding sufficient organizational effort without making the study unreasonably long.
In addition, the code used in the study was drawn from an introductory computer vision course, featuring various visual outputs and complex fine-tuning tasks, which aligned well with our goal of creating complex analytical scenarios. 
The code used in the study is provided in the supplementary material.

\subsection{Procedure and Task} 
\label{sec:user_study_task}

\textbf{Introduction:}
We introduced the consent form, and participants adjusted their chairs and headsets to ensure clear visibility.

\textbf{Training:}
Participants began by reviewing the codes in a Jupyter Notebook on a desktop setup. 
They asked to identify non-linear execution order, such as branches and loops. 
This task involved interpreting the code to uncover how the analytical process was structured, allowing participants to focus solely on the organizational process in the later tasks.
Participants then moved to the immersive environment to familiarize themselves with the provided interactions and reduce the learning curve.
Training concluded once participants demonstrated a clear understanding of the given codes and provided interactions, typically within 15–20 minutes.
While the \guideT{} required longer training, they had already become familiar with the code during the desktop training phase.

\textbf{Task 1: Building execution-focused organization.}
Participants then began the study tasks. 
They were introduced to the immersive environment and informed that the notebook had been imported from Training. 
In this setup, the notebook was displayed in a circular layout with a left-to-right execution flow, resembling a linear sequence.
They were asked to reorganize such linearity into a non-linear structure that better conveyed a complex workflow and supported effective navigation.

\textbf{Task 2: Building narrative-focused organization.}
In the final phase, participants were asked to adjust the organizations to highlight key findings (e.g., particular code cells of interest). 
This task asked participants to transform the notebook into a narrative-focused organization, making the organization accessible to both technical and non-technical audiences, with a focus on visual narrative.
While we specified which cells to highlight, participants had full freedom in making organizational decisions.

\textbf{Questionnaires.}
After completing all tasks, participants completed a Likert-scale survey, which included the System Usability Scale (SUS) and NASA Task Load Index (TLX), and provided qualitative feedback on their experience. 
The entire study lasted approximately 90 minutes and concluded once all questionnaires had been completed.
Our user study followed a full-factorial within-subjects design, counterbalanced using a Latin square.

\subsection{Participants and Apparatus}
We recruited 20 participants (12 male, 8 female, ages 18–35) from a university mailing list with experience in data science, machine learning, and computational notebooks. 
12 participants had used VR, while the rest had no prior experience. 
We used a Meta Quest Pro headset connected to a PC (Intel i7-11800H, NVIDIA RTX 3070) in a large obstacle-free space ($30 m^2$).

\subsection{Measures}
We collected quantitative data and interaction logs for each study condition. 
Specifically, we measured the following: \textbf{Time}: the duration from the start of the organization process until the task completion.
\textbf{Number of Movement}: A recorded sequence of the participant's positions over time.
\textbf{Travel Distance}: The total distance participants traveled to complete the task. We calculate the total physical distance each participant traveled using the Euclidean distances between consecutive recorded positions.

In addition to quantitative data, we collected subjective ratings using a seven-point Likert scale to assess \textbf{physical demand}, \textbf{mental demand}, perceived \textbf{difficulty} in organizations, \textbf{learn a lot} to use, and overall \textbf{preference} for each condition. 
Finally, we also collected qualitative feedback to gain deeper insights into participants’ perceptions and experiences across conditions.

\section{Results}
\label{sec:results}
In this section, we report the quantitative results and qualitative feedback for each condition. 
For dependent variables (or transformed values) that satisfied the normality assumption, we applied linear mixed-effects models to examine the effect of independent variables~\cite{Bates2015}. 
Model comparisons were conducted using log-likelihood ratio tests, and when significant effects were found, we performed Tukey’s HSD post hoc tests for pairwise comparisons based on least-square means~\cite{Lenth2016}. 
We used predicted-versus-residual plots and Q–Q plots to visually inspect homoscedasticity and normality.
For dependent variables that did not meet the normality assumption, we employed the Friedman test to analyze the effect of the independent variable, followed by the Wilcoxon–Nemenyi McDonald–Thompson tests for pairwise contrasts.
Furthermore, for ordinal data, we used the Wilcoxon signed-rank test to conduct within-subject pairwise comparisons~\cite{wilcoxon1945individual}. 
Results are reported with the test statistic ($V$), standardized $Z$ value, and $p$-value. 
We also calculated effect sizes using $r = \tfrac{Z}{\sqrt{N}}$, with $N$ representing the number of participants~\cite{fritz2012effect}.
Statistical significance is reported for: $p < 0.001$ (***), $p < 0.01$ (**), $p < 0.05$ (*), and $p < 0.1$ (·).
Furthermore, we include mean values with 95\% confidence intervals (CI) and report Cohen’s d to indicate effect sizes. 
% Full statistical results are available in the supplementary materials.

\begin{figure}
    \centering
    \includegraphics[width=0.95\columnwidth]{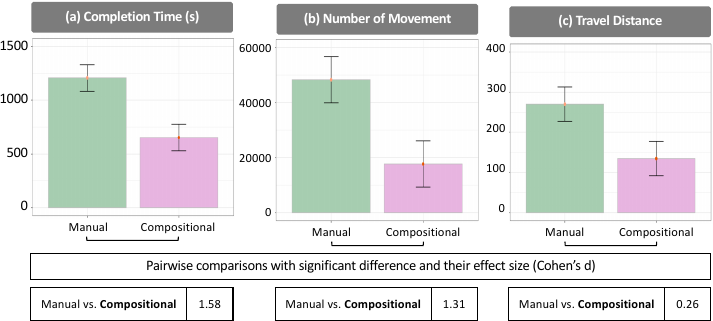}
    \vspace{0.4em}  
    \caption{Participants’ (a) completion time, (b) number of movements, and (c) travel distance across conditions. Solid lines indicate statistically significant differences (p $<$ 0.05), with tables showing effect sizes; bold text indicates the outperforming condition.}     
    \label{fig:quant}
\end{figure}

\subsection{Quantitative Results}
\begin{figure}
    \centering
    \includegraphics[width=0.9\columnwidth]{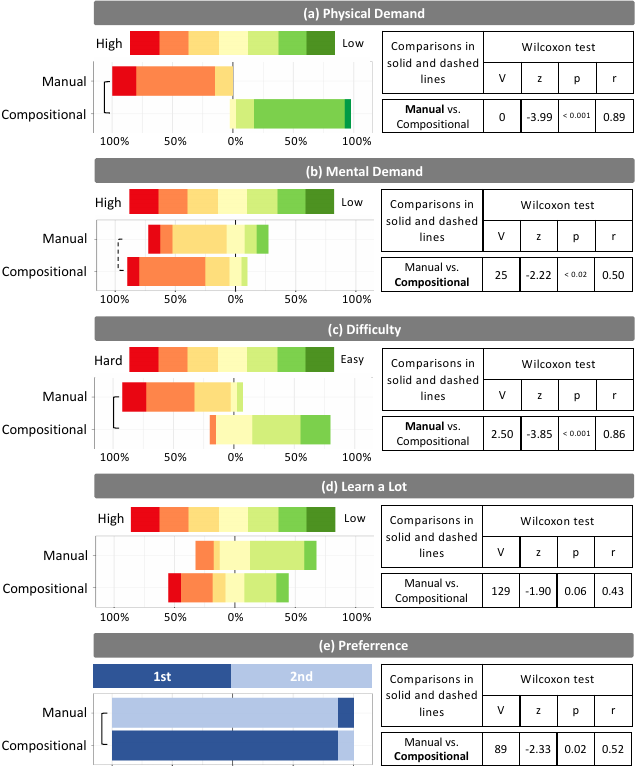}
    \vspace{0.4em}
    \caption{Participants' subjective ratings. (a) physical demand, (b) mental demand, (c) difficulty level, (d) learn a lot, and (e) preference. Solid lines indicate statistical significance with p $<$ 0.01, and dashed lines indicate statistical significance with p $<$ 0.05, with tables showing effect sizes; bold text indicates the outperforming condition.}    
    \label{fig:ratings}
    \vspace{-0.1em}
\end{figure}

The following sections present quantitative results across the conditions, see \autoref{fig:quant} and \autoref{fig:ratings}.

\textbf{Time}: 
Completion time was significantly affected by the \guideT{} (***). 
A linear mixed-effects model showed a significant effect of condition, $\beta = -0.88$, $SE = 0.11$, $\chi^2(1) = 28.38$, $p < 0.0001$. 
Pairwise comparisons also revealed that participants took substantially less time ($\beta = -0.88$, $SE = 0.11$, $z(40) = -7.68$, $p < 0.0001$) in \guideT{}~($avg.~652.95s$, $CI= 104.87s$) than in \manualT{}~($avg.~1206.90s$, $CI= 206.71s$).

\textbf{Number of Movement}:
Participants in \guideT{} moved significantly less than those in \manualT{} (***). 
A linear mixed-effects model showed a significant effect of condition, $\beta = -1.34$, $SE = 0.20$, $\chi^2(1) = 24.15$, $p < 0.0001$. 
Pairwise comparisons also revealed that participants moved substantially less ($\beta = -1.34$, $SE = 0.20$, $z(40) = -6.67$, $p < 0.0001$) in \guideT{}~($avg.~17694.9$, $CI= 5469.68$) than in \manualT{}~($avg.~48301.5$, $CI= 14447.97$).

\textbf{Travel Distance}:
Total physical distance traveled was also significantly lower in \guideT{} (***). 
A linear mixed-effects model showed a significant effect of condition, $\beta = -1.27$, $SE = 0.29$, $\chi^2(1) = 14.15$, $p < 0.0001$. 
Pairwise comparisons also revealed that participants moved substantially less distance ($\beta = -1.27$, $SE = 0.29$, $z(40) = -4.33$, $p < 0.0001$) in \guideT{}~($avg.~131.86$, $CI= 38.85$) than in \manualT{}~($avg.~348.71$, $CI= 142.53$).

\textbf{Ratings.}
Subjective ratings showed significant differences in physical demand (***) and perceived difficulty (***). 
Participants rated physical demand at 5.80 in \manualT{} and 1.95 in \guideT{}.
For perceived difficulty, ratings were 4.80 in \manualT{} and 2.35 in \guideT{}. 
In addition, 19 out of 20 preferred \guideT{} (***). 
However, \guideT{} showed marginal effects (*), with higher ratings for mental demand (2.45 in \manualT{} compared to 3.45 in \guideT{}) and effort to learn (2.70 in \manualT{} compared to 3.65 in \guideT{}) show no significant effects.
In addition, the \guideT{} got 80.18 SUS score, indicating that the system is ``Acceptable'' with ``Excellent'' adjective rating \cite{bangor2008empirical}.

% \textbf{Usability of embodied compositional structuring.}
% Participants significantly favored the \guideT{} with an average of 4.7 out of 5 in impressiveness.
% Additionally, participants reported that they would like to frequently use the system (4.4 out of 5), various functions were well integrated (4.7 out of 5), and organizing a larger number of artifacts was not cumbersome (4.4 out of 5).
% Moreover, 13 participants felt that the system supported a comprehensive set of execution orders.
% Furthermore, 19 participants preferred the compositional structuring to organize artifacts in an immersive computational notebook (see \autoref{fig:ratings}-(e))
% To this end, the \guideT{} got 80.18 SUS score, indicating that the system is ``Acceptable'' with ``Excellent'' adjective rating \cite{bangor2008empirical}.

\subsection{Qualitative Feedback.}
We conducted a qualitative analysis using thematic coding.
Two researchers developed an initial scheme from the first five participants and applied it consistently across all responses. 
This allowed us to extract key insights across conditions. 
Full coding details are provided in the supplemental materials.

% Good
% Flexibility: 10
% Easy Navigation (head rotation): 11 (
% Easy Organization (better than desktop): 5

% Bad
% Physical fatigue: 16
% Hard Organizations: 9
% give up: 7
% Cannot see: 3

\manualB{}\textbf{.}
Participants favored the \textit{ease of navigation} (11) and \textit{flexible workspace management} (10), with some noting the system was \textit{easy to organize} (5).
However, most experienced \textit{physical fatigue} (16) from organizing cells individually, and several \textit{gave up} (7) due to the increasing efforts. 
Others mentioned needing to move closer to view content \textit{not visible from a distance} (3), adding to the \textit{difficulty in maintaining organization} (9).

% Good
% Easy to organize: 19
% Template cover all: 13
% Well-Oranized structures: 11
% Intuitive Interactions: 7

% Bad
% Hard to move big size: 4
% Lot to learn: 7 
% Less Freedom: 5
\guideB{}\textbf{.}
Compared to the \manualT{}, participants found the \guideT{} makes them \textit{easy to organize} (19) and helps them build \textit{well-organized structures} (11) due to its \textit{comprehensive structures} (13).
They also appreciated the \textit{intuitive triggering mechanism} (7). 
However, some noted there was \textit{a lot to learn} (7) the gestures during training, while others cited \textit{reduced flexibility} (5) in adjusting spacing and \textit{difficulty interacting with large structures} (4).

\section{Observations \& Discussions}

% \textbf{Usability of embodied compositional structuring.}
% Participants significantly favored the \guideT{} with an average of 4.7 out of 5 in impressiveness.
% Additionally, participants reported that they would like to frequently use the system (4.4 out of 5), various functions were well integrated (4.7 out of 5), and organizing a larger number of artifacts was not cumbersome (4.4 out of 5).
% Moreover, 13 participants felt that the system supported a comprehensive set of execution orders.
% Furthermore, 19 participants preferred the compositional structuring to organize artifacts in an immersive computational notebook (see \autoref{fig:ratings}-(e))
% To this end, the \guideT{} got 80.18 SUS score, indicating that the system is ``Acceptable'' with ``Excellent'' adjective rating \cite{bangor2008empirical}.

\begin{figure}
    \centering
    \includegraphics[width=0.8\columnwidth]{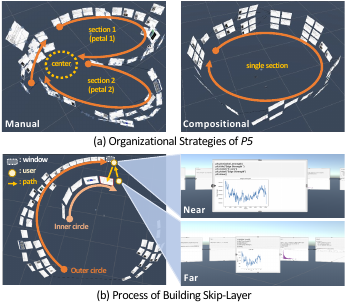}
    \vspace{0.2em}    
    \caption{Illustrations of unique strategies and behaviors during the organization process. (a) A participant (P5) in the \manualT{} created multiple sections from a central point, forming a flower-like shape, with each ``petal'' representing a sub-task in the analysis. (b) While participants were building Skip-Layer, they moved back and forth, as the content was not clearly visible from a distance.} 
    \label{fig:organization}
\end{figure}

\begin{figure*}
    \centering
    \includegraphics[width=1\textwidth]{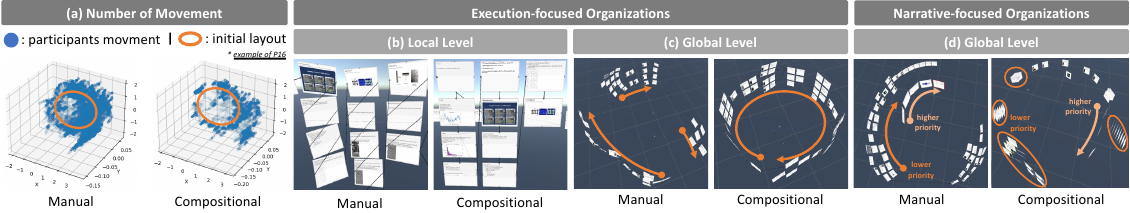}
    % \vspace{0.1em}  
    \caption{Illustrations of comparison between \manualT{} and \guideT{}: (a) Participants moved less in \guideT{}, indicated by fewer blue dots. (b) In execution-focused tasks, \guideT{} enables building structures with clearly placed execution indicators. (c) Participants created more continuous workflows in \guideT{}. (d) In narrative-focused tasks, participants in \guideT{} actively organized lower-priority cells by piling, folding, or hiding execution indicators.} 
    \label{fig:limit_manual}
\end{figure*}

% 1. Time
% 2. execution order (skip process)
% 3. alignment, spacing
% 2. efforts
% 3. Moving cost

% Highlight for execution order (this is the one they did not want to build clear organizations)
% Highlight the (movenet efforts)
\textbf{\guideB{} enabled participants to rapidly build structures with reduced efforts (RQ1).}
We observed that the \guideT{} consistently outperformed \manualT{} in task completion time and reduced effort (see \autoref{fig:quant}-(a) and \autoref{fig:ratings}-(a)). 
In \manualT{}, participants had to manually align cells while managing execution logic, whereas \guideT{} provided predefined structures that reduced the need for fine-tuning. 
At the global level, moving groups of cells and execution indicators in \guideT{} significantly reduced repetitive movements (see \autoref{fig:limit_manual}-(a)).
One participant (P8) explicitly highlighted the effectiveness of the embodied composition framework, noting that they applied the same organizational strategies in both conditions but completed the task 204\% faster with \guideT{}.
Subjective ratings echoed these findings, with 19 participants preferring the \guideT{} for maintaining consistent spacing, alignment, and clarity. 
As one remarked, ``I can organize the workflow very easily and fast (P8).''

% Externalization of intended organizational strategies shown to be better in \guideB{} than \manualB{} (RQ2).
\textbf{Participants still abandoned organizational refinement in \manualB{}, but significantly less in \guideB{} (RQ2).}
Participants in the \manualT{} frequently abandoned both local and global organizational refinements, consistent with prior observations~\cite{in2023table}.
At the local level, 15 participants let execution indicators cross cells, obscuring code and reducing interpretability~\cite{chattopadhyay2020s} (see \autoref{fig:limit_manual}-(b)). 
Globally, 10 participants gave up workspace refinement as the effort required to move multiple artifacts increased (see \autoref{fig:limit_manual}-(c)).
In contrast, fewer participants in \guideT{} abandoned refinements (8 locally, 5 globally). 
Additionally, 14 participants in \manualT{} opted for keeping organizations from execution-focused tasks in a narrative-focused task, whereas most participants in \guideT{} actively folded, clustered their structures, and hid all execution indicators for narrative-focused tasks (see \autoref{fig:limit_manual}-(d)).
We believe that the \guideT{} encourages participants to build intended structures even when the analytical focus shifts.
As one participant noted, ``I just gave up on having nice structures (P14, \manualT{}),'' while another reflected, ``I could organize it as I intended, even nicer shape and alignment (P2, \guideT{}).''

% oversimplify

%%%%%%%%%%%%% Also, talk about the Pile %%%%%%%%%%%%%%%

% It provides effective completion time, but limits the flexiblity
% Unique structure found in manual, but not with guidance.
\textbf{\guideB{} potentially led participants to similar organizational strategies, in contrast to the more diverse strategies observed in \manualB{} (RQ3).}
We observed that participants in \guideT{} tended to adopt similar organizational strategies, particularly at the local-level structures.
Eleven participants transformed long linear sequences into large-sized Multi-row or Multi-column Grid structures. 
In contrast, 15 participants in \manualT{} employed more varied approaches, using multiple smaller Grid structures to fit more compactly. 
At the global level, while differences were less evident, one notable example was participant P5, who used a flower-like multiple-section workspace in \manualT{} but adopted a single-section workspace in \guideT{}, consistent with most others. 
These findings partially align with prior work indicating that external assistance can discourage users from conducting additional analysis and may reduce the diversity of approaches to creative problem-solving~\cite{vasconcelos2023explanations}.

% They rather pull one by one
% Because Far - Near proxy
% Include the Pile!
% Participants frequently utilized the compositions to build Skip-Pile, but rarely used them for Skip-Layer (RQ4).
\textbf{Participants manually built Skip-Layer structure in \\ \guide{} (RQ4).}
During narrative-focused tasks, participants frequently adopted the Skip-Layer structure to highlight key findings, but only four used the gestures to build. 
Despite positive feedback during training, most built Skip-Layers manually. 
Participants often walked closer to inspect cells (near) and then positioned key items into distant locations (far). 
We believe this behavior is caused by visibility issues---code and outputs were difficult to see from afar (see \autoref{fig:organization}-(b)).
As one participant noted, ``I cannot clearly see the code from a far distance (P1).''
This suggests that the push-and-pull gesture was ineffective.
Therefore, we conclude that gestures for interacting with distant content, especially when readability is critical, require further refinement.

\textbf{Gesture-based triggering mechanisms yielded mixed results when applied to a larger set of interactions (RQ4).}
Gesture-based interaction provides an intuitive way for performing operations~\cite{cordeil2019iatk, yang_pattern_2022}.
Similarly, we found that the provided triggering mechanisms were perceived as intuitive by 7 participants, with one participant (P1) explicitly stating that the Proxy Window helped them perform gestural interactions more clearly, as it provided visual feedback of their actions.
However, an equal number (7) found the gestures hard to learn and recall (see \autoref{fig:ratings}-(d)). 
Despite extensive training, all participants requested gesture reminders during tasks. 
As one participant remarked, ``It was hard to remember every operation (P13).''
This likely contributed to increased mental demand (see \autoref{fig:ratings}-(b)), as they needed to remember 13 distinct interactions.
Therefore, we recognize that gesture-based interaction places a mental burden on users when they are required to remember a large number of distinct gestures.

\textbf{Limitations\&Future Work.}
While the embodied composition framework improved organization speed and reduced effort, it also indicated limitations in the triggering mechanism.
Potential improvement could be integrating Focus+Context lenses~\cite{yang_pattern_2022}, widely used in visualization to support simultaneous overview+detail~\cite{chimera1992value}.
We believe this would not only address challenges related to inspecting code from a distance but also minimize physical movements to walk closer to see the contents in large workspaces.
Additionally, participants struggled to recall lot number of gestures due to the absence of a visible reference, unlike WIMP interfaces.
To address this, we envision a hybrid approach combining gesture and WIMP interactions: common or user-defined operations could be gesture-based~\cite{wobbrock2009user}, while others remain accessible via menus, balancing intuitiveness and scalability.

\section{Conclusion}
\label{sec:conclusion}

In this study, we introduced an embodied composition framework to support the organization process within immersive computational notebooks. 
Our findings show that it significantly reduced both the time and effort required to organize content, enabling participants to maintain clear execution logic, proper alignment, and spacing.
However, the embodied composition framework also presents trade-offs. 
While it streamlines the organizational process, it may oversimplify the complex workflow as analysts rely heavily on the composition framework, potentially limiting opportunities for expressing complex analytical reasoning.
In summary, we believe that the embodied composition framework offers potential for supporting the organizations of immersive computational notebooks to improve the usability of immersive workspaces.

%\section*{Acknowledgments}
% This work was supported in part by NSF I/UCRC CNS-1822080 via the NSF Center for Space, High-performance, and Resilient Computing (SHREC).

\begin{acks}
This research was supported by industry, government, and institute members of the NSF SHREC Center, which was founded in the IUCRC program of the National Science Foundation, and NSF grant IIS-2441310.
\end{acks}

\bibliographystyle{ACM-Reference-Format}
% \balance
\bibliography{main}

\end{document}